\documentclass[twocolumn,showpacs,preprintnumbers,amsmath,amssymb]{revtex4}

\usepackage{graphicx}
\usepackage{dcolumn}
\usepackage{bm}
\usepackage{textcomp}

\begin{document}

\preprint{APS/123-QED}

\title{Influence of the inhomogeneous field at the tip \\ on quantitative piezoresponse force microscopy}

\author{Tobias Jungk}
\author{\'{A}kos Hoffmann}
\author{Elisabeth Soergel}

\email{soergel@uni-bonn.de}

\affiliation{Institute of Physics, University of Bonn,
Wegelerstra\ss e 8, 53115 Bonn, Germany}

\date{\today}

\begin{abstract}
Ferroelectric domain imaging with piezoresponse force microscopy
(PFM) relies on the converse piezoelectric effect: a voltage applied
to the sample leads to mechanical deformations. In case of PFM one
electrode is realized by the tip, therefore generating a strongly
inhomogeneous electric field distribution inside the sample which
reaches values up to $10^8\,$V/m directly underneath the apex of the
tip. Although often assumed, this high electric field does not lead
to an enhancement of the piezoelectric deformation of the sample. On
the contrary, internal clamping of the material reduces the observed
deformation compared to the theoretically expected value which
depends only on the voltage thus being independent of the exact
field distribution.
\end{abstract}


\maketitle


Ferroelectric domain patterns are the basis of a multitude of
applications such as quasi-phase-matched frequency converters
\cite{Fej92}, electro-optic scanners \cite{Gah01}, nonlinear
photonic crystals \cite{Bro00}, and ultra-high density data storage
devices \cite{Cho05}. For further improvement of the domain patterns
their visualization with high lateral resolution is indispensable.
Although several techniques have been developed \cite{Soe05},
piezoresponse force microscopy (PFM) has become a standard technique
in recent years mainly because of its easy use. However, the
interpretation of the obtained images is still challenging,
therefore quantitative data is published very rarely. This
deficiency is often justified by the presumption that due to the
strong dependency of the electric field on the tip radius, which in
general is not known exactly, a quantitative analysis of the data is
not possible. Arguing that way, however, ignores the fact that, at
least in a first approximation, not the electric field distribution
but only the applied voltage determines the piezoelectric
deformation of the sample. Although this statement is self-evident
from theoretical considerations, we carried out experiments with
different single-domain crystals, comparing the measured deformation
underneath the tip with and without an additional top electrode.

PFM is based on the deformation of the sample due to the converse
piezoelectric effect. The piezoresponse force microscope is a
scanning force microscope (SFM) operated in contact mode with an
additional alternating voltage applied to the tip. In piezoelectric
samples this voltage causes thickness changes and therefore
vibrations of the surface which lead to oscillations of the
cantilever that can be read out with a lock-in amplifier. In
ferroelectric samples different orientations of the polar axis of
adjacent domains lead to a domain contrast, i.\,e., the domain faces
are displayed as bright or dark areas in PFM images (an overview of
the PFM technique can be found in \cite{Alexe}). The generally
observed frequency dependence of those measurements
\cite{Agr05,Scr05,Lab00} was recently be explained by a
system-inherent background \cite{Jun05a}. We also proposed a
detection scheme that allows a straight forward quantitative
analysis of the obtained data \cite{Jun05b}. In this contribution we
investigate the influence of the strongly inhomogeneous electric
field of the tip on the piezoelectric deformation measured with PFM.

The (longitudinal) converse piezoelectric effect says that in an
external electric field $E$ a piezoelectric material of thickness
$t$ undergoes a thickness change $\Delta t$ proportional to the
appropriate piezoelectric coefficient $d$:
\begin{equation}
\label{01}
\frac{\Delta t}{t} = E\;d \quad \Rightarrow \quad  \Delta t =  U \,d
\end{equation}
Note that the thickness change $\Delta t$ does not depend on the
thickness $t$ of the sample but only on the applied voltage $U$.
That is why for piezoelectric transducer elements, not thick
crystals but stacks of multiple piezoelectric discs, all of them
connected individually, are utilized \cite{piezo}.

In case of PFM the electric field is generated by the tip and a
large homogeneous back electrode. Typical dimensions are: tip radius
$r\leq 100$\,nm, thickness of the crystal several 100\,\textmu m,
and back electrode some mm$^2$. The voltage applied between tip and
back electrode usually is 10 -- 20\,V$_{\rm pp}$ which leads to
electric field strength at the very tip of the order of $E= U/r =
10^8\,$V/m (for $r=100$\,nm and $U=10$\,V).
Assume a coordinate system such that the $z$-axis is perpendicular
to the surface of the sample under investigation (thus $z$ lies in
the symmetry axis of the tip).
Because the electric field $E$ has a radial symmetry,
piezomechanical deformations due to field components perpendicular
to $z$ cancel out at the position of the tip apex. The electric
field $E$ relevant for PFM measurements can therefore be reduced to
$E_{\rm z}$.
Equation~(\ref{01}) is also valid if the electric field $E_{\rm z}$
is not homogeneous along the sample thickness $t$ because $\int_0^t
\vec E\,{\rm d}\vec s = U$. Therefore the inhomogeneous distribution
of the electric field $E_{\rm z}$ underneath the tip has no effect
on the piezoelectric deformation along the $z$ direction.

\begin{figure}
\includegraphics{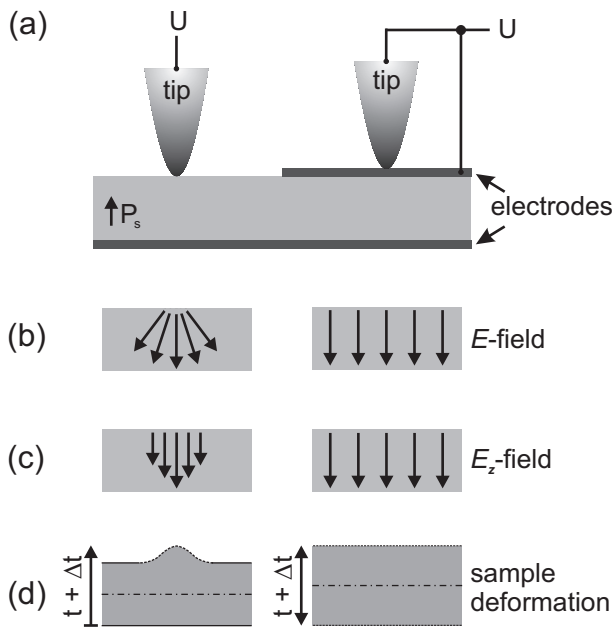}
\caption{\label{fig:Jungk1}
(a) Setup for the determination of the influence of the electrode
configuration on the piezoelectric deformation when measured with
PFM. (b) Electric field distribution inside the crystal and (c) its
component along the $z$-direction. (d) The resulting piezoelectric
deformation of the sample. $P_s$:~spontaneous polarization,
$U$:~applied voltage, $t$:~sample thickness, $\Delta t$:~thickness
change}
\end{figure}

In order to support this statement we prepared samples of different
single crystals where part of the topface was covered with a
conductive, 50\,nm thick gold-layer of 10\,mm$^2$ area, the backside
being fully covered with a gold layer to ensure a homogeneous back
electrode (Fig.~\ref{fig:Jungk1}(a)). In case of the PFM tip acting
as electrode the field distribution is strongly inhomogeneous
whereas with the large gold electrode a homogeneous field inside the
crystal is generated (b). In (c) the relevant electric field
component $E_{\rm z}$ is depicted. The expected deformations are
schematically shown in (d). In case of a homogeneous field (right
side) the whole volume of the sample undergoes a thickness change
whereas in case of the tip acting as an electrode, the deformation
occurs only in a restricted volume at the very surface of the
crystal.

For PFM measurements a frequency of 10 - 100\,kHz usually is chosen
for the alternating voltage applied to the tip. With the tip acting
as electrode, the generally observed frequency dependence can be
attributed to mechanical resonances of the SFM head, why for
different samples the same frequency spectra are observed.
Accounting for this background one can obtain quantitative,
frequency independent results \cite{Jun05a}. When using a
homogeneous electrode to apply the voltage, however, the whole
sample is vibrating, thereby also exciting the sample holder. In
this case the frequency spectrum is no longer independent neither on
the sample used nor on its mounting. To overcome this difficulty, we
carried out measurements at very low frequencies (some Hz) thereby
avoiding any mechanical excitation both of the SFM head, and of the
sample and sample holder.

Tabular~\ref{tab1} summarizes the experimental results obtained with
single crystals of lithium niobate ($\rm LiNbO_3$), $\alpha$-quartz
as well as a periodically poled potassium titanyl phosphate (KTP)
\cite{Can06}. For all samples, the values determined by PFM with the
tip acting as electrode are smaller by roughly a factor of three
compared to the data obtained with homogeneous electrodes. For
$d_{11}$ of $\alpha$-quartz there is agreement for the published
values of the piezoelectric coefficients. For $\rm LiNbO_3$,
however, the values reported for $d_{33}$ range from 6~--~7\,pm/V
\cite{War68,Smi71,Oka89} to 16~--~23\,pm/V \cite{Yam67,Chk71,Bha05}.
Interestingly for the other piezoelectric coefficients $d_{15}$,
$d_{22}$, and $d_{31}$ there is a reasonable good consensus in those
publications. As the KTP is a multidomain sample it is not possible
to obtain meaningful data with a large top electrode (see also
\cite{Jun06a}). The goal of this contribution, however, is not to
discuss possible reasons for these surprising discrepancies. From
our measurements we tend to join the groups having reported the
large values.

\begin{table}
\caption{\label{tab1}Comparison of the piezoelectric deformation
measured with PFM (tip as electrode or extra top electrode) for
three different samples ($d_{33}$~for~$\rm LiNbO_3$, $d_{33}$ for
KTP, and $d_{11}$  for $\alpha$-quartz). In addition values from
literature with the corresponding references are given.}
\begin{ruledtabular}
\begin{tabular}{lccccc}
Sample & PFM-tip & PFM-top & Literature & Ref.\\
       & [pm/V]  & [pm/V]  & [pm/V] &     \\
       &&&&&\\
$\rm LiNbO_3$ & 5 - 7 & 19 - 22  & 16 - 23& \cite{Yam67,Chk71,Bha05}\\
&&& 6 - 7 & \cite{War68,Smi71,Oka89} \\
KTP & 6 & -- &  10 - 26 & \cite{Gra97}\\
$\alpha$-quartz & 0.8 & 2.3 & 2.3 & \cite{Bec58}\\
\end{tabular}
\end{ruledtabular}
\end{table}

So why are the values measured with the tip acting as electrode
too small by a factor of three? We suppose that mainly internal
clamping of the sample constrains the mechanical deformation.
Because of the inhomogeneity of the electric field, the full
deformation occur in a very small volume ($\mu$m-sized). This,
however, leads to clamping and thus to a reduced piezoelectric
deformation. Note that the load of the tip can not cause this
effect, because the crystal undergoes the same load underneath the
tip also with a large electrode. Increasing the load by a factor
of 100 does not affect the values measured either. Furthermore we
carried out a series of measurements with different cantilevers,
varying both spring constant and tip radius, but could not observe
any dependence.

The strongly inhomogeneous field at the very tip might also lead to
another effect reducing the deformation of the crystal: it is not
ensured that the material still answers linearly applying an
electric field of $10^8$\,V/m, as it is present at the very tip.
This would also lead to values for the piezoelectric coefficient
smaller than expected.

A further point is: why are the values measured with the tip as
electrode to small by roughly the same factor for all three samples?
We explain this by the stiffness of those materials which is of the
same order of magnitude \cite{War68,Smi71,Bec58}, the samples
thereby undergo similar mechanical clamping. Performing experiments
with single crystals exhibiting very different elastic properties
should therefore lead to other factors. Ceramics, however, behave
completely different why they are not suited for comparative
measurements.

The results as described above sustain the statement declared in the
beginning of this contribution that the electric field distribution
at the very tip does not lead to an enhancement of the piezoelectric
deformation measured with PFM.

There is an important point in the technical realization of the
experiment described above: When measuring the piezoelectric
deformation with PFM and a large top electrode, it is absolutely
mandatory to connect the top electrode directly because applying the
voltage $U$ to it only via the tip generally fails. This is also the
case for fresh gold layers and metal coated tips due to the small
contact area and the inevitable surface adsorbates when measuring
under ambient conditions. In case of no electrical contact, however,
the top electrode acts as a metallic shielding, thus no
piezoelectric deformation but only the system inherent background is
observed \cite{Jun05a,Jun06a}. Note that because the crystal is
supported by a stable mount, also in the case of a homogeneous
electrode the whole thickness change $\Delta t$ is measured as
thickness change at the top face.

Calculations like the one in the paper by Kolosov et
al.~\cite{Kol95} are simply wrong: admittedly, the electric field at
the very tip can reach values up to $E_{\rm tip} = 2.5 \times
10^7$\,V/m, however, this electric field persists not along the
whole thickness of the crystal. Therefore the value of $\Delta t =
2\,t\,E_{\rm tip}\,d_{33} = 30\,$nm (sample thickness $t=0.3$\,mm
and $d_{33}=2\times 10^{-12}\,$m/V) is incorrect.

In this contributions we tried to clarify, despite its simplicity,
an apparently still not generally accepted feature of PFM imaging:
the surface deformation of the sample is not enhanced due to the
high electric field at the very tip. We hope having disposed the
idea of large PFM signals due to electric field enhancement of the
tip. This is not the case - neither from a theoretical point of view
nor from experimental results. In contrary, the values measured with
PFM are smaller than expected.

\begin{acknowledgments}

Financial support of the DFG research unit 557 and of the Deutsche
Telekom AG is gratefully acknowleged.

\end{acknowledgments}

\end{document}